\documentclass[showpacs,twocolumn, nofootinbib]{revtex4}
\usepackage{amsmath,amssymb,amsthm}
\usepackage{graphicx}
\usepackage{mathrsfs}

\newcommand{\dr}{\rightarrow}

\def\jj{{\cal J}}

\def\mm{{\cal M}}

\newcommand{\mn}{{\mu\nu}}

\newcommand{\beq}{\begin{equation}}
\newcommand{\eeq}{\end{equation}}

\begin{document}
\title{Planck-Scale Modified Dispersion Relations and  Finsler Geometry}
\author{F.~Girelli\footnote{girelli@sissa.it}, S.~Liberati\footnote{liberati@sissa.it},  L.~Sindoni\footnote{sindoni@sissa.it} } \affiliation{SISSA,
Via Beirut 2-4, 34014 Trieste, Italy  and INFN, Sezione di Trieste}

\begin{abstract}
A common feature of all Quantum Gravity (QG) phenomenology approaches is to consider a modification of the mass shell condition of the relativistic particle to take into account quantum gravitational effects. The framework for such approaches is therefore usually set up in the cotangent bundle (phase space). However it was recently proposed that this phenomenology could be associated with an energy dependent geometry that has been coined ``rainbow metric". We show here that the latter actually corresponds to a Finsler geometry, the natural generalization of Riemannian geometry. We provide in this way a new and rigorous framework to study the geometrical structure possibly arising in the semiclassical regime of QG. We further investigate the symmetries in this new context and discuss their role in alternative scenarios like Lorentz violation in emergent spacetimes or Deformed Special Relativity-like models.
\end{abstract}

\maketitle

\section{Introduction}
The quest for a quantum theory of gravity (QG) has nowadays a long standing record that stretches for several decades in the past. While this research has lead to great achievements in theoretical physics the final answer still remains hidden to us. In particular the lack of observational and experimental tests has been a constant problem in the development of QG models. Fortunately in the last decades the improved sensitivity of our experiments and observations has lead to unexpected opportunities to tests sub-Planckian consequences of several QG scenarios. Among these opportunities let us remember deviations from Newton's law at very short distances \cite{Hoyle:2004cw, Hoyle:2000cv}, Planck-scale fuzziness of spacetime \cite{Amelino-Camelia:1999gg}, possible production in TeV-scale QG scenarios of mini-black holes at colliders \cite{Dimopoulos:2001hw} or in cosmic rays~\cite{Anchordoqui:2001cg}, QG induced violations of discrete symmetries of the Standard Model \cite{Ellis:1995xd,Kostelecky:2005mj} as well as spacetime symmetries \cite{Garay:1994en}. This broad field of research goes under the general name of quantum gravity phenomenology.

It should not come as a surprise that a preeminent role in this field of research has been played by possible tests of QG scenarios proposing departures from Lorentz invariance (LI) at energies approaching the Planck scale (for an extensive review see {\em e.g.}~\cite{Mattingly:2005re}). While LI is deeply rooted both in quantum field theory as well as in General Relativity, still it is strictly linked to the idea that spacetime would have the same continuous structure at any energy scale: large boosts naturally uncover the structure of spacetime at arbitrary small scales and it is unclear how this could be conciliated with the existence of a fundamental scale for the quantum gravitational phenomena, {\em i.e.}~the Planck scale.

Departures from standard Lorentz invariance have nowadays been proposed within several QG scenarios. A succinct list includes: arguments based on string field theory tensor VEVs~\cite{KS89}, cosmologically varying moduli scenarios~\cite{Damour:1994zq}, spacetime foam models~\cite{Amelino-Camelia:1997gz}, semiclassical spin-network calculations in Loop QG~\cite{Gambini:1998it,loopqg}, as well as non-commutative geometry studies~\cite{Hayakawa,Mocioiu:2000ip,Carroll:2001ws,Anisimov:2001zc} and some brane-world scenarios~\cite{Burgess:2002tb}. Albeit none of the  above cited calculations can be considered at the moment a conclusive evidence that departures from Lorentz invariance are generic in QG models, they can indeed be considered a robust hint that Lorentz violations could be a theoretical possibility in most of the scenarios we are envisaging for QG. Moreover they generally agree in predicting departures from LI in the form of modified dispersion relations (MDR) for elementary particles of the form
\begin{eqnarray}
E^2&=&m^2+p^2  +D(p,\mu,M)
\nonumber\\
&=& m^2+p^2+ \sum_{n=1}^\infty \alpha_{n}(\mu,M)  \; p^n,
\label{eq:mod-disp}
\end{eqnarray}
where $p=\sqrt{||\vec p||^2}$, $\alpha_n$ are dimensional coefficients, $\mu$ is some particle physics mass scale, and $M$ is the scale associated to the new physics responsible for the correction to the dispersion relation  (which is typically taken to be the Planck mass: $M_{\rm P}\approx 10^{19}$ GeV).

While it is not the scope of the present paper to explain how constraints can be cast on such MDR (see {\em e.g.}~\cite{Jacobson:2005bg}) it is interesting to note that these are generally requiring to choose a well defined dynamical framework. While the most stringent constraints have been so far obtained considering the MDR as a by-product of an effective field theory with Planck suppressed Lorentz violating operators~\cite{Jacobson:2005bg}, there is an alternative point of view that tries to preserve the relativity principle and extend it to the case in which there are two (the speed of light and the Planck energy) invariant scales in place of a single one (the speed of light). This proposal goes under the name of Doubly or Deformed Special Relativity (DSR) \cite{Amelino-Camelia:2000mn}. Unfortunately while on the  one hand we have precise formulations of  DSR in momentum  space on   the   other   hand,   the
implementation of DSR in spacetime is a more subtle subject and it is a theme of intense debate at present time \cite{Amelino-Camelia:2000mn, Magueijo:2001cr, Magueijo:2002am, Judes:2002bw,Kimberly:2003hp}.  This is a crucial point as it is clear that no DSR proposal could be really competitive with Lorentz violation scenarios, lacking a concrete understanding of the spacetime symmetries associated with this deformed Lorentz symmetry and their possible geometrical interpretation.

In this direction, a very well studied scenario for the spacetime realization of DSR is non-commutative geometry (see {\em e.g.}~\cite{Amelino-Camelia:2000mn, Kowalski-Glikman:2002jr}).
While some promising results linking DSR to some special form of spacetime non-commutativity  have been found in 2+1 dimensions~ \cite{Freidel:2003sp}, we are still lacking a consistent physical  picture in 3+1 dimensions. Moreover attempts to develop a quantum field theory associated with different forms of non-commutativity, a much needed step in order to be able to effectively cast phenomenological constraints, led to
highly non-trivial quantum field theories (possibly with problematic features as the IR/UV-mixing~\cite{Minwalla:1999px, Matusis:2000jf, Amelino-Camelia:2002au}). 

Our investigation here will be based on different assumptions: we shall not presume any manifold non-commutativity but show instead that MDRs of the sort shown in (\ref{eq:mod-disp}) can be interpreted as due to a different form of departure from standard spacetime structure: its pseudo-Riemaniann geometry. 
In doing this we shall not assume neither that such departures from the standard spacetime structure are due to UV Lorentz violations or to DSR.  

Attempts to reconstruct a spacetime structure starting from a MDR have been already made in the existent literature (see {\em e.g.}~\cite{Kimberly:2003hp}). 
Interestingly the key feature of the corresponding spacetime is to be described by an energy dependent metric, called a rainbow metric \cite{Magueijo:2002xx}.
This is a natural concept from the QG point of view and arose in different contexts like spacetime foam~\cite{Ellis:1999uh}, the Renormalization Group (RG) applied to gravity \cite{Girelli:2006sc} or as a consequence of averaging over QG fluctuations~\cite{Aloisio:2005qt}. Perhaps even more intriguing is the fact that we have concrete realization of rainbow geometries in the so called analogue models of gravity where phonons at high energy show exactly MDR like (\ref{eq:mod-disp}) as a consequence of the energy dependence of the effective geometry induced by the condensed matter background structure (see {\em e.g.}~\cite{Barcelo:2005fc}).

However the rainbow metric idea proposed in \cite{Magueijo:2002xx} still lacks a rigorous formulation since it involves  a metric defined on the tangent bundle, while depending on a quantity associated to the cotangent bundle
({\em i.e.}~the energy)\footnote{Note that the aforementioned examples of effective rainbow metrics do not have this problem since the energy scale is either associated to an
external quantity (momentum exchange between the string and the
topological defect) in the case of the string Liouville theory, or the
theory is defined only in the cotangent space in
the RG approach~ \cite{Girelli:2006sc}.}. Henceforth the aim of the present paper will consist in defining the geometry associated to a MDR in a mathematically rigorous way.

In searching for a geometrical structure leading to dispersion relations of the form  (\ref{eq:mod-disp}) it is clear that we will be forced to abandon the safe harbor of Riemannian geometry as they certainly cannot accommodate any energy dependence\footnote{Remarkably the same intuition has been used 35 years ago  to bypass the GZK cutoff \cite{Kirzhnits:1972sg}.
Unfortunately the authors have only considered a very restricted set of modified dispersion relations (\textit{i.e.}~homogeneous of degree two in momentum) which was erroneously supposed to lead to new physics at high energy.}.
In this sense it can be illuminating to look again at the situations appearing in analogue models. In fact there it can be showed that departures from exact LI at low energies can be naturally described via the considerations of the so called Finsler geometries~\cite{Barcelo:2001cp, Weinfurtner:2006wt}. Finsler structures are the most studied generalizations of Riemannian geometry and are defined starting from norms on the tangent bundle instead than from inner products~\cite{baochernshen}.

In what follows we shall further investigate this possibility and show that indeed {\em any} MDR of the kind (\ref{eq:mod-disp}) can be seen as describing the propagation of a particle on a Finsler geometry.  In this sense we prove that any rainbow metric is indeed a Finsler metric. While we cannot show that this definitely hints towards a deformed symmetry scenarios (with respect to a Lorentz violation one) we hope that this will help further developments by providing a well posed and consistent framework within which one can hope to set up a proper quantum field theory (as needed to cast constraints).

The plan of the paper will be the following: after a first section devoted to Finsler geometries, we shall deal directly with the way one can reconstruct a Finsler structure from a MDR. We shall then consider the symmetries associated to this new structures making a distinction between those relevant only on the manifold and those defined on the whole (co)tangent bundle. Finally we shall discuss the relevance of our results highlighting some possible developments of our investigation.

\section{Finsler Geometry}

Let us start with a brief review of the basic notions relevant for Finsler geometries.
(Further material can be found, for example, in \cite{baochernshen, rund}.)
Finsler geometry is a generalization of Riemannian geometry: instead
of defining an inner product structure over the tangent bundle, we
define a norm $F$. This norm will be a real function $F(x,v)$ of a
spacetime point $x$ and of a tangent vector $v\in T_x M$, such that
it satisfies the usual norm properties namely
\begin{itemize}
  \item $F(x,v)\neq 0$ if $ v \neq \mathbf{0}$,
  \item $F(x,\lambda v)= |\lambda | F(x,v), \qquad \lambda \in \mathbb{R}$.
\end{itemize}
 From the norm squared
$F^2(x,v)$ we can define the so called Finsler metric:
\begin{equation}\label{eq:finslermetric}
g_\mn(x,v) = \frac{1}{2} \frac{\partial^2 F^2}{\partial v^\mu \partial
v^\nu},
\end{equation}
which we require to be continuous and non-degenerate.  Using
Euler's theorem\footnote{If $Z(v)$ is a homogenous  function of
degree $r$, then $v^i\frac{\partial Z}{dv^i}= rZ(v)$.} on
homogeneous functions, it can be shown that \eqref{eq:finslermetric}
is equivalent to \beq F(x,v) = \sqrt{g_\mn(x,v)v^\mu v^\nu}. \eeq
This shows that $g_\mn(x,v)$ is a homogeneous function of degree
zero of the vector $v$.  Also, since by definition $g_\mn$ is non degenerate, it admits an inverse $g^\mn$  such that $ g^\mn(x,v)g_{\nu\alpha}(x,v)={\delta^{\mu}}_\alpha$.

Given the above definition a few comments are in order.
First of all it is clear that  Riemannian geometry can now be seen as the special
case of Finsler geometry in which $F$ is given by an inner product ({\em i.e.}~when $g_\mn$ does not depend on $v$). It is also important to
stress that the Finsler metric defined in \eqref{eq:finslermetric}
is defined in the Euclidean signature, that is when $g_\mn$ is
positive definite. The extension to the pseudo-finslerian case,
although not always obvious, can be done in many situations (see {\em e.g.}~the appendix of \cite{Weinfurtner:2006wt}).

Finally, since the Finsler metric is homogenous of degree zero,  it
cannot be defined on the zero section of the tangent bundle. Finsler
geometry is therefore usually defined over the slit-tangent bundle,
which is simply the tangent bundle minus the zero section
\cite{baochernshen}: \beq \breve{T}M = \bigcup_{x\in M} (T_{x}M
\setminus \{{\bf 0}\}). \eeq This is only a mathematical subtlety,
since smooth curves are defined to have non-vanishing tangent
vector, and we will never encounter explicitly this problem in our
discussion.

We have given in the definition of a Finsler norm the condition of
homogeneity,
$F(\lambda v)=|\lambda|F(v)$,
according to the usual postulates regarding norms over vector spaces. However there are
situations in which it is good to have an asymmetry with respect to
the inversion $v\rightarrow -v$. In this case, we restrict the
homogeneity condition to the so called \textit{positive homogeneity}
condition:
$F(\lambda v) = \lambda F(v) , \lambda >0$.
Notice that this provides an extension of the usual definition of norms over vector spaces.

While in Riemannian geometry the inverse metric was directly inducing a
scalar product between forms, to do the same thing in Finsler
geometry one must use some caution. The duality between vectors and
forms which is used in Riemannian geometry is given by the formula:
\beq \omega_{\mu}(v)= g_{\mu\nu}(x) v^{\nu}. \eeq
Due to the fact that the metric tensor is represented by a non degenerate matrix, we can
invert this relation to express a vector in terms of its dual form
\beq v^{\mu}(\omega) = g^{\mu\nu}(x)\omega_{\nu}. \eeq

Therefore, given the scalar product between vectors we can naturally
induce a scalar product between forms exploiting this duality. More
precisely, we put: \beq \langle \omega_1,\omega_2 \rangle_{forms}
\equiv \langle v(\omega_{1}),v(\omega_2) \rangle_{vectors}. \eeq

We have to extend this formalism to the case of Finsler metrics
since we want to deal with the four momentum of a particle, which is
more easily treated as a form.

We define the dual form to a vector by the equation: \beq
\omega_{\mu} = g_{\mu\nu}(x,v)v^{\nu}. \eeq In terms of the norm this
relation can be written as: \beq \omega(v)_{\mu} = \frac{1}{2}\frac{\partial F^2(x,v)}{\partial v^{\mu}}. \eeq It is
important to note that, if $g$ is a non degenerate Finsler metric, then the map just defined
 between forms and vectors is invertible. We
can exploit this inverse mapping to define the norm of a form given
the norm of a vector:

\beq G(x,\omega) = F(x,v(\omega)). \eeq

The tensor obtained from this norm plays the same role of the
inverse metric tensor in Riemannian geometry, and it is simply given by \beq
h^{\mu\nu}(x,\omega) = \frac{1}{2} \frac{\partial ^2
G^2(x,\omega)}{\partial{\omega}_{\mu}\partial \omega_{\nu}}.\eeq

We can connect this tensor to the inverse metric
$g^{\mu\nu}(x,v)$ just using the definition of $G$. \beq
h^{\mu\nu}(x,\omega) = g^{\mu\nu}(x,v(\omega)). \eeq

 The above derivation concludes our formal definition of the Finsler geometries and shows that these geometrical structures define a proper working framework which can be used indifferently either in the tangent bundle or in the cotangent bundle. However, in order to complete our treatment of Finsler geometries, a brief discussion about the treatment of curvature in these spaces is due.

The theory of curvature of Finsler spaces is more involved than the case of Riemannian geometry. Nevertheless it is worth to give some key
ideas for understanding the possibilities given by this kind of
structure to describe new gravitational physics. If one considers
the case of a position dependent Finsler norm $F(x,\dot x)$, and looks at the
geodesic equation, one obtains:
\begin{equation}
    \ddot{x}^{\mu} + \Gamma^{\mu}_{\nu\rho}(x,\dot{x})\,\dot{x}^{\nu}\dot{x}^{\rho}
    =0,
    \label{geoeq}
\end{equation}
where the Christoffel's symbols  $ \Gamma^{\mu}_{\nu\rho}(x,\dot{x})$ contain derivatives of the metric  only in the coordinates, while keeping an explicit dependence on the velocity (See the appendix for an explicit derivation of $\Gamma^{\mu}_{\nu\rho}(x,\dot{x})$ and further discussion).
Without going into mathematical details, we can hence say that not only the metric structure
of the theory is velocity dependent, but also curvature effects, like tidal forces, can have modifications due to the nonlinear nature of the connection.

We define the Finsler metric to be ``flat" when there exists a {\em global} coordinate system in which the connection coefficients vanish. Given that
the connection $\Gamma^{\mu}_{\nu\rho}(x,\dot{x})$ contain derivatives of the metric only with respect to the coordinates, the above definition is equivalent to the statement that the ``flat'' Finsler metric is globally independent on the coordinates. Note however, that such  metric will still  be velocity dependent  so it will not be the trivial Minkowski one.

The above discussion shows that the concept of Finsler structure allows  to treat rigorously the idea of
velocity/momentum dependent metric. Consequently, we can fit into
this scheme the idea that there is a microscopic structure of
spacetime which cannot be described by ordinary (pseudo-)Riemannian
geometry. Notice, however, that the extension we are considering is
still metric, \textit{i.e.} we are not introducing non-metric structures as
in metric-affine theories of gravity \cite{Hehl:1976kj, Hehl:1994ue,Sotiriou:2006qn}.

\section{Modified dispersion relations and Finsler norms}
 A particle moving in a Finsler manifold is described by the action:
\beq
\label{eq:actionfinslerparticle}I = m \int_a ^b F(x,\dot{x})\,
d\tau ,
\eeq
where the norm $F(x,\dot{x})$ can in principle also depend  on several parameters like for example the mass of the particle and the Planck scale.

Action (\ref{eq:actionfinslerparticle}) is simply the straightforward generalization to Finsler geometry of the standard action for the free relativistic particle
\beq
I = m\int_{a}^{b}
\sqrt{g_{\mu\nu}(x)\dot{x}^{\mu}\dot{x}^{\nu}}\,d\tau,
\eeq
and as such allows to recover Eq.~(\ref{geoeq}) as the Euler--Lagrange equation and to define the canonical momentum
\beq
p_{\mu} =m \frac{\partial F}{\partial \dot{x}^{\mu}}=m \frac{g_{\mu\nu}(x,\dot{x}) \dot{x}^\nu}{F},
\label{mom}
\eeq
which satisfies the generalized mass-shell condition
\begin{eqnarray}
h^{\mu\nu}(x,p) p_{\mu}p_{\nu} &=& m^2
g^{\mu\nu}(x,\dot{x})\frac{\textstyle g_{\mu\rho}(x,\dot{x})\dot{x}^{\rho}g_{\nu\sigma}(x,\dot{x})\dot{x}^{\sigma}}
{\textstyle g_{\alpha\beta}(x,\dot{x})\dot{x}^{\alpha}\dot{x}^{\beta}}\nonumber\\
&=&m^2.
\end{eqnarray}

It is important to note that, due to the homogeneity of the norm in its vector argument, the action (\ref{eq:actionfinslerparticle}) is reparametrization invariant. If instead we have a positively homogenous norm, the action will then be only invariant under time ``Orientation Preserving" (OP) reparametrizations. This more general set of actions will be called OP reparametrization invariant.

We now want to show that for any MDR of the form (\ref{eq:mod-disp}) it is possible to introduce a (OP) reparametrization invariant  action whose Lagrangian can be identified with a (positively homogenous) Finsler norm. Also, given that we are interested in modifications of Special Relativity, we shall consider from now on the case of a ``flat" Finsler norm $F(x,\dot{x})=F(\dot{x})$.
Let us stress that we will always assume through all the paper that coordinates and momenta satisfy the canonical Poisson bracket $\{x^\mu , p_\nu \}={\delta^\mu}_\nu$.

Let us start noticing that the assumed reparametrization invariance of the action implies that the Hamiltonian  $H=\dot{x}^\mu p_\mu-\cal{L}$ is identically zero.  Consequently in order to implement a given mass-shell condition $\mm(p)=m^2$ (like for
example \eqref{eq:mod-disp}) we have to introduce a Lagrange multiplier $\lambda$ so that  the action becomes
\begin{equation}\label{eq:fundamaction}
    I = \int \left(\dot{x}^{\mu}p_{\mu} -
    \lambda(\mm(p)-m^2)\right)\, d\tau.
\end{equation}
Note that the reparametrization invariance of this action implies that $\mathcal{M}(p) = \mathcal{M}(-p)$ should hold.

To find out what is the connection between the action \eqref{eq:fundamaction} and a Finsler
geometry we have to pass to the Lagrangian formalism. To do this we
use one of the Hamilton's equations: \beq \dot{x}^{\mu} = \lambda
\frac{\partial \mm}{\partial p_{\mu}}. \eeq

In the case in which this relation is invertible we can
express $p$ as a function of $\dot{x}$ and $\lambda$, hence  obtaining an action of the form:
\beq
I= \int \mathcal{L}(\dot{x},\lambda)d\tau.
\eeq
We can eliminate $\lambda$ from this action just using the equation of motion obtained from varying the action with respect to this parameter so that ${\cal L}(\dot{x},\lambda)\to{\cal L}(\dot{x},\lambda(\dot{x}))$.

So, in the end, we derive a Lagrangian which is a function only of $\dot{x}$ and that is  homogeneous of degree one, consequently it can be identified with a Finsler norm.
\begin{equation}
{\cal L}(\dot{x},\lambda(\dot{x}))=mF(\dot{x}).
\end{equation}
Note however that if $\mathcal{M}(p) \neq \mathcal{M}(-p)$ the action is only OP reparametrization invariant and the Lagrangian is a positively homogeneous Finsler norm.

To make this discussion clearer, let us consider the simple
example of a 
MDR for a
particle moving in two dimensions
\begin{equation}
\mathcal{M}(p_0,p_1) = p_0 ^2 - p_1 ^2 - \frac{\alpha}{M} p_1 ^3,
\label{eq:mdr-ex}
\end{equation}
where $\alpha$ is a dimensionless quantity.
The corresponding action is given by\footnote{In the following we will use the notation $x=(t,\mathrm{x})$.}
\begin{equation}
I=\int \left( \dot{t}p_0 + \dot{\mathrm{x}}p_1 - \lambda \left( \mathcal{M} -
m^2  \right) \right) d\tau.
\end{equation}
Clearly this action is only OP reparametrization invariant, 
since the mass shell condition is not invariant under time inversion.

Let us consider the Legendre transform to the Lagrangian formalism.
We obtain:
\begin{eqnarray}
\dot{\mathrm{x}} & =& \lambda \left\{ \mathrm{x}, \mathcal{M} \right\} = \lambda \left(
-2 p_1 - \frac{3\alpha}{M}p_1 ^2 \right),\label{tizio}\\
\dot{t} &=&2 \lambda p_0\label{caio}.
\end{eqnarray}
Performing the Legendre transformation, we find that the Lagrangian
describing the motion of the particle is given by:
\begin{equation}
\mathcal{L} = \frac{1}{4} \frac{\dot{t}^2 - \dot{\mathrm{x}}^2}{\lambda} -
\frac{1}{8}\frac{\alpha}{M} \frac{\dot{\mathrm{x}}^3}{\lambda^2}+\lambda m^2+O(\alpha^2).
\label{ex-lagr}
\end{equation}
The equation of motion obtained varying $\lambda$ in the above Lagrangian is rather difficult to solve but we can find an approximate solution for $\lambda$ by requiring that $\alpha p_1 \ll M$ so that for any momenta well below the Planck scale we can safely  keep only the order $\alpha$ corrections.
This procedure gives the following (approximate) solution:
\begin{equation}
\lambda(\dot{x}) \approx \frac{\sqrt{\dot{t}^2-\dot{\mathrm{x}}^2}}{2m}
- \frac{\alpha}{2M} \frac{\dot{\mathrm{x}}^3}{\dot{t}^2 -
\dot{\mathrm{x}}^2}.
\label{ex-lam}
\end{equation}
Using this relation we obtain the particle Lagrangian:
\begin{equation}
\mathcal{L}(\dot{x},\lambda(\dot{x}))=m\sqrt{\dot{t}^2-\dot{\mathrm{x}}^2} - \frac{\alpha m^2}{2M}
\frac{\dot{\mathrm{x}}^3}{\dot{t}^2 - \dot{\mathrm{x}}^2},
\label{ex-plagr}
\end{equation}
which allows us to identify the Finsler norm
\begin{equation}
F_m(\dot{x})=\sqrt{\dot{t}^2-\dot{\mathrm{x}}^2} - \frac{\alpha m}{2M}
\frac{\dot{\mathrm{x}}^3}{\dot{t}^2 - \dot{\mathrm{x}}^2}.
\label{ex-norm}
\end{equation}
 One can easily check that, as expected,
this line element is a positively homogeneous Finsler norm. When $\alpha=0$ we recover the Lagrangian for the relativistic
particle.

In the case of more general mass shell functions $\mm(p)$, these
passages can be more involved due to algebraic complications, but
the final result must be the same: \textit{ if the action describing the
propagation of a particle is  (OP) reparametrization  invariant  then,
whatever is the mass-shell condition\footnote{The massless case should be treated
with care, since the relation between the
multiplier $\lambda$ and the mass might not have a smooth limit for
$m\rightarrow 0$. In general one cannot expect then, to be able to remove the multiplier from the action (at least in the way we did in the  massive case).  However, this is a problem as well in the case of the massless particle in Special Relativity, where one cannot eliminate $\lambda$ from the action since this parameter is undetermined by the equations of motion. On the other hand, determining  the trajectories of the massless particle with a modified dispersion relation can  be done in the same way as in Special Relativity.}, the trajectories followed by this particle are the geodesics of a particular, possibly particle dependent, (positively homogenous) Finsler metric, given by the procedure just described.}

A final comment concerns the fact that even assuming a universal coefficient $\alpha$ in (\ref{eq:mdr-ex}) still the MDR corresponds to a Finsler norm which is mass dependent: particles with different masses see different Finsler structures. This is consistent with the fact that Finsler
norms have no scale embedded in them (by homogeneity), while in general modified dispersion relations can contain several energy scales at which different physical effects are turned on.

If on the contrary the mass shell condition
involves a single universal metric $g^{\mu\nu}(p)$ homogeneous of degree zero in momentum
\begin{equation}
    \mm(p)=g^{\mu\nu}(p)p_{\mu}p_{\nu}
    \label{trF},
\end{equation}
then different particles with different masses propagate over the same Finsler
structure which is given by the Legendre transform of such a metric in tangent space.
It has to be noticed however that in this case the homogeneity of the metric plus its uniqueness\footnote{These are  precisely the assumptions  of \cite{Kirzhnits:1972sg}.} imply that there is no interesting phenomenology associated to the MDR. If instead one allows for a particle dependence of the metric in (\ref{trF}) --- but still preserving homogeneity in  momentum --- then one would recover (in the limit $p \gg m$) dispersion relations of the kind $E^2=p^2+m_i^2+\eta_i p^2$ for which a rich new phenomenology and consequently constraints are expected  (as for example \cite{Coleman:1998ti}).

\section{Symmetries}
The effect of modifying the special relativistic dispersion relation
is twofold: on one side we have a modification in the relation
between energy, momentum and mass. On the other, we have a
modification of the symmetry group of the the system, and
correspondingly a modification in the associated Noether charges.

In order to give a physical meaning to our mathematical formalism we
have to give precise definitions of physical observables, and in
particular of conserved charges. For particles moving along
geodesics of pseudo-Riemannian manifolds we can use Killing vector
fields to generate conserved quantities

\begin{equation}
    Q = \xi^{\mu}(x)\dot{x}^\nu g_{\mu\nu}(x),
\end{equation}
where $\xi$ is a Killing vector field for the metric $g$ and
$x(\tau)$ is an affine parametrization of the geodesic. For example,
the energy of a particle can be defined as the projection of the
speed along the direction of the Killing vector field associated to
time translations.

It is clear that we need to provide a definition of Killing vector
field of a Finsler metric. This can be done \cite{rund} (although we
shall not give here the mathematical derivation for sake of conciseness) just by applying the definition
of Lie derivative to the Finsler metric tensor, taking into account
that there is a contribution to the derivative coming from the
dependence of the metric on the tangent vector on which it is
evaluated.

As we have discussed in the previous section, the natural way of describing the motion of particles with modified dispersion relation is Finsler geometry, which can be formulated in an elegant way in the (co)tangent bundle. Therefore, when considering the issue of symmetries, we have to consider all the transformations on the (co)tangent bundle which leave invariant the Finsler metric. As a consequence, the diffeomorphisms of the manifold are only a special case for these symmetries.

For Killing vector fields which generate these manifold diffeomorphisms,
we have the Killing equations:
\begin{widetext}
\begin{equation}
\label{eq:killing}
    g_{\rho \mu}(x,\dot{x}) \partial_{\nu} \xi^{\rho} + g_{\nu\rho}(x,\dot{x})
    \partial_{\mu}\xi^{\rho} + 2 C_{\rho\mu\nu}(x,\dot{x}) \frac{\partial \xi^{\rho}}{\partial
    x^{\sigma}} \dot{x}^{\sigma} + \frac{\partial g_{\mu\nu}}{\partial
    x^{\rho}} \xi^{\rho} =0,
\end{equation}
\end{widetext}
where we have defined the Cartan tensor:
\begin{equation}
    C_{\mu\nu\rho} = \frac{1}{2}\frac{\partial g_{\mu\nu}(x,\dot{x}) }{\partial
    \dot{x}^{\rho}}.
\end{equation}
This equation is obtained exactly as the Killing equation in Riemannian geometry, just considering the Lie derivative of the metric tensor. Using equation (\ref{eq:killing}), it is now easy to show that if $\xi$ is a Killing vector field, the associated
Noether charge can be written as
\begin{equation}\label{eq:conservedcharge}
    Q= \xi^{\mu}(x) \dot{x}^{\nu} g_{\mu\nu}(x,\dot{x}).
\end{equation}
In this way we can provide a consistent definition of energy and
momentum relating them to the Noether charges.\\

In particular, for the specific case of flat Finsler geometries ({\em i.e.}~a metric tensor
depending only on the velocity and not on the position), we see that
spacetime translations are symmetries of the system and hence one
can apply equation (\ref{eq:conservedcharge}) to this special case. Therefore
the conserved quantities associated to spacetime translations are:
\begin{equation}
    \mathcal{E} = g_{0\mu}(\dot{x}) \dot{x}^{\mu},\quad \mathcal{P}_i  =   g_{i\mu}(\dot{x})\dot{x}^{\mu}.
\end{equation}
On shell, using an affine parameter for the geodesic and modulo a mass rescaling, these
quantities coincide with the canonical momenta appearing in the
action \eqref{eq:fundamaction}, or, equivalently, in the action
\eqref{eq:actionfinslerparticle} given that
\begin{equation}
    p_{\mu} = m \frac{\partial F}{\partial \dot{x}} = \frac{m}{F}
    g_{\mu\nu}(\dot{x}) \dot{x}^{\nu},
\end{equation}
and that, in the case of an affine parametrization, F is constant along
a geodesic. Hence, taking into account normalizations,
the energy momentum co-vector obtained using Noether's theorem obeys
the mass shell condition appearing in \eqref{eq:fundamaction}.
Therefore the momenta appearing in the MDR are really the physical momenta associated to
spacetime translations, so that $p_\mu\leftrightarrow\partial_\mu$.

This discussion shows how we can treat the case of diffeomorphisms of the manifold
with the same language used in Riemannian geometry. However, as we have stressed, these are only
a subset of the transformations of the full tangent bundle. To show that there are other transformations which
are relevant, we now pass to the discussion of the Lorentz group and its action for the case of a modified dispersion
relation.
Since using Killing vector fields
can be difficult, we will describe the
symmetries of a MDR using the Hamiltonian formalism. The infinitesimal generators
$\cal G$ of the symmetry group obey the equation
$$\{{\cal G},\mm(p)\}=0.$$ By construction the translation generators given by $p_\mu$
leave invariant the MDR $$\{p_\mu,\mm(p)\}=0.$$ This is consistent
with the fact that the system described by the action
\eqref{eq:actionfinslerparticle} is invariant under translations. On
the other hand the usual Lorentz generators $J_\mn= x_\mu
p_\nu-x_\nu p_\mu$ are not commuting anymore with the MDR
$$\{J_\mn,\mm(p)\}\neq0.$$ In particular, there could be a situation in which
while the generators of rotations are conserved, the generators of
boosts do not commute with the MDR.

With respect to the fate of the Lorentz group, one could adopt two
different points of view.  The first one consists in considering the MDR as a manifestation of
a \textit{symmetry breaking} mechanism which destroys the original
Lorentz symmetry, reducing it to a smaller group. This could be related to the fact that the underlying theory,
as in the case of analogue models \cite{Barcelo:2005fc}, might not have the Lorentz group as fundamental symmetry group, this latter appearing only as an approximate symmetry in the low energy regime, leading to our geometric description of spacetime in terms of a pseudo-Riemannian manifold.

The alternative point of view is to look for a deformed action of the
Lorentz group, i.e. a representation of the Lorentz group acting on
the whole phase space {\em i.e.}~mixing configuration and momentum space. In this case it is
therefore necessary to deform the Lorentz generators, for example
adding extra terms
\begin{equation} J_\mn \dr \jj_\mn = J_\mn + \alpha_i C^i_\mn(x,p,M),\end{equation}
where $C^i_\mn(x,p,M)$ are functions on phase space to be
determined by the equations
\begin{eqnarray}
& \{\jj_\mn,\mm(p)\} = 0,\nonumber\\
& \\
& \left\{ \jj_{\mu\nu}, \jj_{\rho\sigma} \right\} = \eta_{\mu\sigma} \jj_{\rho \nu} + \eta_{\nu\rho} \jj_{\sigma\mu} + \eta_{\mu\rho} \jj_{\nu\sigma} + \eta_{\nu\sigma} \jj_{\mu\rho}.\nonumber
\end{eqnarray}

Here the $\eta_{\mu\nu}$ are the components of the Minkowski metric,
which are the structure constants of the Lorentz group, while our
physical metric tensor is given by the Finsler metric tensor
obtained from the norm. The action of the deformed Lorentz
infinitesimal generators on phase space coordinates is given by
\begin{widetext}
\begin{eqnarray}\label{symTTM}
&&\{x^\beta, \jj_\mn\}=\{x^\beta, J_\mn\}+ \alpha_i \frac{\partial}{
\partial
p_\beta}C^i_\mn(x,p,M)=x_\mu\delta_\nu^\beta-x_\nu\delta_\mu^\beta +
\alpha_i \frac{\partial}{\partial p_\beta}C^i_\mn(x,p,M),\\&&
\{p_\beta, \jj_\mn\}= \{p_\beta, J_\mn\}- \alpha_i
\frac{\partial}{\partial
x^\beta}C^i_\mn(x,p,M)=-p_\mu\delta_\nu^\beta+p_\nu\delta_\mu^\beta -
\alpha_i \frac{\partial}{ \partial x^\beta}C^i_\mn(x,p,M).
\end{eqnarray}
\end{widetext}
This should be compared with the special relativistic counterpart.
In the case of Minkowski spacetime, the Lorentz group acts on the
position and on momenta separately. They do not mix.
In the general Finsler case we expect a mixing, because the functions $C^i_\mn$
can be very complicated and not just of the form $xp$. This means
that in certain cases we can expect to find a representation of the
Lorentz group acting on the phase space in such a complicated way.
Making the Legendre transformation we can translate this statement
by saying that a symmetry transformation for a Finsler line element
can be a map acting on the tangent bundle. This map, in general, is
not the lift to the tangent bundle of a symmetry acting only on the
manifold as in Riemannian geometry. This is expected in Finsler
geometry since, in some sense, it is the geometry of the tangent
bundle \cite{lagrangegeom}.

It should be noted that the deformation of the Lorentz group implies in general non canonical commutation relations between boost and momenta, affecting  therefore the definition of the Poincar\'e group.  One should check with care the consequences of this fact in the context of
particle physics, in which the role of the Poincar\'e group is
crucial in determining the vacuum state and the classification of
particles.

On the other hand, one could try to complete this deformed Lorentz group with deformed
momenta in such a way to recover the usual Poincar\'e group. To establish which of these groups is the physical one, the
deformed Poincar\'e group or the remnant from the breaking of the Lorentz group, we would have to do, as usual,  quantum field theory tests by looking at processes which are able to discriminate between the two scenarios.

\textit{It is important to recall that, while the action of the symmetry group is unique in phase space, given a specific MDR, the realization in the tangent bundle of this group is particle dependent, due to the explicit mass dependence of the Finsler metric. Therefore we cannot give in the tangent bundle a unique, absolute, spacetime representation of the deformed Lorentz group.}\\

\section{Conclusion}
To summarize, we have showed that the notion of Finsler metric is the geometric structure encoding the notion of rainbow metric. More precisely we showed how  a metric in the cotangent space constructed from a MDR is related to a (positively homogenous) Finsler norm in tangent space, the key point of the analysis being the time (OP) reparametrization invariance.

The Finsler metric arising from the MDR is then particle dependent: each particle of different mass will see a different Finsler metric. We could have expected this from  the intuition behind the rainbow metric:  in the cotangent space, the metric is  dependent on the momentum of the particle, so that in the tangent space, the metric should depend both on velocity and mass. A metric depending on velocity is a Finsler metric, a rainbow metric is then naturally a mass dependent Finsler metric. Note that as a direct consequence, in the case of many particles we obtain  a multi Finsler metric structure. This is similar to the construction that appears in the Coleman-Glashow analysis \cite{Coleman:1998ti}.

The discussion made in this paper shows that there could be a regime in
which the standard geometrical description breaks down in a serious way.
Not only we have to pass from Riemannian geometry to Finsler geometry, but
we have that each particle sees a different geometrical structure depending on its mass, making
the notion of ``the spacetime geometry'' or even of ``tangent bundle geometry'' completely meaningless.

This should be compared with the case of analogue models for gravity.
There, the concept of a Lorentzian spacetime geometry is a low-energy,
emergent one, while the true underlying spacetime theory is completely
different. The fact that the notion of spacetime geometry is shaking as it
stands is just a manifestation of something deep: the MDR are a consequence of the fact that we are probing the microstructure of spacetime,
which of course could be non-geometric.

However, there could be other ways in which a geometric description could be preserved.
In fact there are good reasons to think that behind the Standard Model there is a unified field theory (GUT) of gauge
interactions. In these theories, fermions are organized in multiplets
within which they share the same features. Originally they are massless,
acquiring a mass only via a Higgs mechanism at suitable energies. If we
postulate, as it seems reasonable, that all fermions are massless when we
start probing quantum gravity effects, then it makes sense to speak about
a geometry (at least in the tangent bundle), since they all have the same MDR in this limit.

Clearly the low energy regime brings into the game the details of the
pattern of the various symmetry breakings, leading to deviations from a
single geometrical picture. Nevertheless, it is also true that at low energies quantum gravity effects are presumably subdominant, so that the Lorentzian spacetime
geometry becomes, to a very good degree of approximation, the
appropriate description for the stage of the Standard Model dynamics.

Alternatively one can remove the mass dependence from (\ref{ex-norm})  by just substituting $\alpha\,m/M$ by a new dimensionless coefficient $\alpha^\prime$ and assuming it to be universal ({\em i.e.}~particle independent). In this case compatibility with observations would however require $\alpha^\prime \ll 1$ hinting that it would have to be a small ratio of some particle physics mass/energy scale, ${\cal M }$, and, presumably, the Planck one: $\alpha^\prime=\mu/M$. Let us notice however that in our example (\ref{eq:mdr-ex}) the above cited rescaling would lead  to a dispersion relation with a LIV term $(\mu/m)  \, p_1^3/M$. This implies that, at least for the case of electrons with a cubic term in the MDR,  this new energy scale would have to be smaller than the electron mass in order to be compatible with current constraints (if EFT can be applied then the best constraints for the dimensionless coefficient of the $p^3/M$ term range nowadays between $10^{-2}$ and $10^{-7}$, see {\em e.g.}~\cite{Jacobson:2005bg}). It is tantalizing that such line of reasoning seems to suggest that $\mu$ should be identified with some sort of IR fundamental scale for gravitational physics complementary to the UV one represented by the Planck mass.

The study of symmetries associated to the Finsler metric showed that there are two possibilities. We can have broken Lorentz symmetries   if we consider that a symmetry transformation should be (as usual) a transformation that lives separately in the manifold and in the (co)tangent space. If we relax this, that is allow for symmetry transformation mixing (co)tangent space and manifold, then the Lorentz symmetries are still present. They are then called {\it deformed}. This construction arises for example in Deformed Special Relativity ({\it e.g.}  in the bicrossproduct basis or DSR1 \cite{Majid:1994cy}). From the geometrical perspective this is a change of paradigm: the fundamental object is not the manifold anymore but the full tangent bundle. This point of view is consistent with the DSR intuition: when trying to implement some universal momentum scale, we are led to unify the notion of spacetime with momentum just as in Special Relativity one unifies space and time by introducing  a universal  speed. Once again this picture is  still consistent with the rainbow metric since to have full information  about a system in this framework, we need to know both its position and momentum.

In this sense the Finsler structure   seems to be the natural framework for a geometric interpretation of DSR. The fact that DSR structures involve usually non trivial symplectic forms does not contradict this. Indeed it is always possible to make a Darboux transformation to obtain a trivial symplectic form \cite{Girelli:2005dc} and perform there our analysis. Alternatively  brute force calculations can also be made as in \cite{Ghosh:2006cb}. Note {\it en passant} that the Darboux transformation implies in general that the new configuration coordinates should be function of both the old configuration and momentum variables, so that on the geometric point of view we are really living on the full tangent bundle.

In conclusion, this new geometric point of view provides a new angle of attack to deal with some of the problems met in the context of Lorentz violations or deformed symmetries as in DSR. For example it might provide new insights for the construction of Effective Field Theory in these contexts: something necessary to make meaningful experimental predictions.
The proposed framework provides also a concrete mathematical set up to describe the QG corrections to the notion of uniformly accelerated observer, horizons... These are exciting questions that we leave  for further studies.

\bigskip

\acknowledgments{}
SL and LS wish to thank S.~Sonego and M.~Visser for illuminating discussions.

\appendix
\section{Geodesic equation}
\label{ap:geodesic}
Let us consider in detail the geodesic equation for Finsler manifolds. We start from the action \eqref{eq:actionfinslerparticle}, with the most general case of position dependent Finsler norm. The Euler--Lagrange equations of motion reduce to:
\begin{widetext}
\beq
\frac{d}{d\tau}\left[ \frac{1}{F} \left( g_{\mu\nu} \dot{x}^{\nu} +\frac{1}{2} \frac{\partial g_{\alpha\beta}}{\partial \dot{x}^{\mu}}\dot{x}^{\alpha}\dot{x}^{\beta} \right) \right]
 - \frac{1}{2}\frac{1}{F} \frac{\partial g_{\alpha\beta}}{\partial {x}^{\mu}}\dot{x}^{\alpha}\dot{x}^{\beta}=0.
\eeq
\end{widetext}
If we introduce the Cartan tensor
\beq
C_{\mu\nu\rho} = \frac{1}{4} \frac{\partial^3}{\partial \dot{x}^{\mu}\partial \dot{x}^{\nu} \partial \dot{x}^{\rho}} F^2 = \frac{1}{2} \frac{\partial}{\partial \dot{x}^{\mu}}g_{\nu\rho},
\eeq
we see that, due to the zero degree of homogeneity of the metric tensor, as a consequence of Euler's theorem on homogeneous functions:
\beq
z \frac{d f}{d z} = n f(z) \Longleftrightarrow f(a z) = a^{n} f(z),
\eeq
applied to the $n=0$ case we obtain:
\beq
\label{eq:fundamentalcartan}
C_{\mu\nu\rho}(x,\dot{x})\dot{x}^{\mu} =C_{\mu\nu\rho}(x,\dot{x})\dot{x}^{\nu}=C_{\mu\nu\rho}(x,\dot{x})\dot{x}^{\rho}=0 .
\eeq
The Cartan tensor is a useful tensorial quantity which can describe the non-Riemannian curvature of the Finsler metric tensor, being non-vanishing even in what we call the ``flat" case.
As a consequence of (\ref{eq:fundamentalcartan}), we obtain that the geodesic equation, in the affine parametrization, becomes
\begin{equation}
\ddot{x}^{\mu} + \Gamma^{\mu}_{\nu\rho}(x,\dot{x})\,\dot{x}^{\nu}\dot{x}^\rho=0,
\end{equation}
where these generalization of the Christoffel's coefficients to the case of Finsler structures are given by the usual
expression:
\begin{widetext}
\begin{equation}
    \Gamma^{\mu}_{\nu\rho}(x,\dot{x}) =   \frac{1}{2}g^{\mu\sigma}(x,\dot{x})\left[-\partial_{\sigma}g_{\nu\rho}(x,\dot{x})
    +\partial_{\nu}g_{\rho\sigma}(x,\dot{x})+\partial_{\rho}g_{\sigma\nu}(x,\dot{x})\right].
\end{equation}
\end{widetext}
We note that these coefficients depend on the velocity as well, and in a non-linear way. This means that to treat the theory of curvature in Finsler space we have to use the language of non-linear connections, for which we refer to the literature \cite{rund,baochernshen,lagrangegeom}.
\bibliographystyle{unsrt}

\end{document}